\documentclass[twocolumn,usenatbib]{mnras}

\usepackage{graphicx}
\usepackage{amsmath}

\def\pmb#1{\setbox0=\hbox{#1}%
    \kern-.025em\copy0\kern-\wd0
    \kern.05em\copy0\kern-\wd0
    \kern-.025em\raise.0433em\box0}

\def\ltsima{$\; \buildrel < \over \sim \;$}
\def\gtsima{$\; \buildrel > \over \sim \;$}
\def\simlt{\lower.5ex\hbox{\ltsima}}
\def\simgt{\lower.5ex\hbox{\gtsima}}
\def\p2Y{\;_2Y}
\def\m2Y{\;_{-2}Y}

\def\mk2{\mu {\rm K}^2}
\def\Planck{\it Planck\rm}

\def\LCDM{$\Lambda$CDM}

\newcommand{\camspec}{{\tt CamSpec}}
\newcommand{\plik}{{\tt Plik}}
\newcommand{\commander}{{\tt Commander}}
\newcommand{\simall}{{\tt SimAll}}

\def\pmb#1{\setbox0=\hbox{#1}%
     \kern-.025em\copy0\kern-\wd0
     \kern.05em\copy0\kern-\wd0
     \kern-.025em\raise.0433em\box0}

\begin{document}

\title[Evidence for a spatially flat Universe]{The evidence for a spatially flat Universe}

\author[George Efstathiou and Steven Gratton]{George Efstathiou and Steven Gratton \\
 Kavli Institute for Cosmology Cambridge and 
Institute of Astronomy, Madingley Road, Cambridge, CB3 OHA.}

\maketitle

\begin{abstract} 
We revisit the observational constraints on spatial curvature
following recent claims that the \Planck\ data favour a closed
Universe. We use a new and statistically powerful \Planck\ likelihood
to show that the \Planck\ temperature and polarization spectra are consistent with a spatially flat
Universe, though because of a geometrical degeneracy 
cosmic microwave background anisotropy spectra on their own do not lead to tight
constraints on the curvature density parameter $\Omega_{\rm K}$. 
When combined with other astrophysical data, particularly
geometrical measurements of baryon acoustic oscillations, the Universe
is constrained to be spatially flat to extremely high precision, with
$\Omega_{\rm K} = 0.0004 \pm 0.0018$ in agreement with the 2018
results of the \Planck\ team. In the context of inflationary
cosmology, the observations offer strong support for models of
inflation with a large number of e-foldings and disfavour models of
incomplete inflation.

\vskip 0.3 truein

\end{abstract}


\section{Introduction}

One of the strongest motivations for the
theory of inflation comes from the observation that our
Universe is very nearly spatially flat. As Guth and others 
\citep{Guth:1981, Linde:1990} have pointed out, exponential expansion during an inflationary
phase provides an elegant solution to the `flatness' problem. As a consequence of exponential expansion,  $\Omega_K = 0$
is a powerful late time attractor.  

The inflationary prediction of a spatially flat Universe is strongly supported by observations.   The  2018 cosmic microwave background (CMB) 
results reported by the \Planck\ team
 (combining \Planck\ temperature and polarization data 
with \Planck\ lensing and baryon acoustic oscillation measurements) give
\begin{equation}
     \Omega_K = 0.0007 \pm 0.0019,  \label{Omega_constraint}
\end{equation}
\citep[][hereafter PCP18]{Params:2018} suggesting that our Universe is
spatially flat to a $1\sigma$ accuracy of $0.2\%$. In addition, the
\Planck\ power spectra are extremely well fit by a nearly scale
invariant adiabatic fluctuation spectrum. Taken together, these two
results offer strong support for the inflationary model.

Recently, however, three papers \citep{Park:2019, DiValentino:2019, Handley:2019}
have presented a different interpretation. These papers point out that
the \Planck\ temperature and polarization data show a preference for a
closed universe (as noted in PCP18). \cite{DiValentino:2019} conclude that the
\Planck\ results favour positive curvature at  the $3.4\sigma$  level (i.e.
a probability to exceed (pte) of $0.034\%$).
 They interpret this high significance level as
evidence for either undetected systematics in the \Planck\ data, new physics,
or an unusual statistical fluctuation (or some combination of all three). \cite{Park:2019} and \cite{Handley:2019} reaches
qualitatively similar conclusions. Since the limit on spatial curvature is of
such fundamental importance to cosmology, we revisit this problem in
this paper.

\section{Curvature and choice of prior}

Let us assume that the spatial curvature of the Universe, $ \Omega_{\rm K} =  K  /(aH)^2$, is of order unity at
the start of inflation (where $a$ is the scale factor and $H$ is the Hubble parameter). If the Universe undergoes
$N$ e-foldings of inflation (ending at $a=a_I$) the curvature parameter at the present day ($a=a_0$) will be
\begin{equation}
         \vert \Omega_{\rm K} \vert = {\rm e}^{-2N} \left [ {(a_I H_I)^2 \over (a_0 H_0)^2 }\right ] . \label{inf1}
\end{equation}
The term in square brackets depends on the duration of the reheating
phase at the end of inflation and the energy scale of inflation and so
is uncertain. For plausible parameters, with an energy scale of
inflation of order $V_I \sim 10^{16}{\rm GeV}$, $(a_I H_I)^2/(a_0
H_0)^2 = {\rm e}^{2N_*}$ with $N_* \approx 60$\footnote{$N_*$ could
  be much lower if the energy scale of inflation is low, but its exact
  value is unimportant for our purposes} \citep[e.g.][]{Liddle:2003}.
A solution of the flatness and horizon problems requires $N >
N_*$. For many models of inflation, {\it e.g.} $\alpha$-attractors
\citep{Carrasco:2018}, the number of e-foldings can be much greater
than $N_*$, in which case our Universe is expected to be spatially
flat to high accuracy. If, however, the number of e-foldings is
comparable to $N_*$, spatial curvature may be detectable today. In any
model involving a small number of e-foldings it is essential that
fluctuations on the curvature scale remain small. This is quite
natural in models of open inflation invoking a Coleman-de Luccia
instanton \citep{Coleman:1980}.  Models of this type have been
discussed by many authors in the past \citep[e.g.][]{Gott:1982,
  Linde:1995, Bucher:1995a, Linde:1999} and they have received renewed
interest in the context of false vacuum decay within a string
landscape \citep[e.g.][]{Freivogel:2006,Yamuchi:2011}. It is also
possible, with moderate fine tunings, to construct models with
positive spatial curvature \citep[e.g.][]{Gratton:2002, Linde:2003}. The exact mechanism is
unimportant for our purposes and neither is the choice of measure.  We will simply 
assume that inflation generates a finite number of e-foldings
with $N>N_*$ skewed to low values:
\begin{equation}
p(N)dN \propto N^{-\alpha}dN, \qquad N>N_*, \label{prob1}
\end{equation}
with $\alpha>1$. Assuming $\vert K \vert/(aH)^2=1$ at the start of inflation, the distribution of
spatial curvatures at the present day is
\begin{equation}
p(\Omega_{\rm K})d\Omega_{\rm K}   = {(\alpha-1) \over 4}  {N_*^{(\alpha-1)} \over (N_* - {1 \over 2} \ln \vert \Omega_{\rm K} \vert )^\alpha}
{d\Omega_{\rm K} \over \Omega_{\rm K}}.  \label{prob2}
\end{equation}
This function is  peaked at $\Omega_{\rm K}=0$, but has tails extending to non-zero values of $\Omega_{\rm K}$. In fact, for the distribution (\ref{prob2}) the probability of finding $\vert \Omega_{\rm K}\vert  > \vert \Omega^*_{\rm K} \vert$ is
\begin{equation}
P ( \vert \Omega_{\rm K}\vert  > \vert \Omega^*_{\rm K} \vert ) \approx {(\alpha -1) \over 2} {\ln \vert \Omega^*_{\rm K} \vert \over N_*},  \qquad {\rm if } \ -{1 \over 2} \ln \vert \Omega^*_{\rm K} \vert  \ll  N_*, \label{prob3}
\end{equation}
and is non-negligible even though the most probable value is $\vert
\Omega_{\rm K} \vert \ll \vert \Omega^*_{\rm K} \vert$. A specific
model of this type of incomplete inflation has been discussed by
\cite{Freivogel:2006}, though these authors used anthropic arguments
in place of $N_*$ to cut off the distribution
(\ref{prob1}). We can therefore view experimental bounds on $\Omega_{\rm K}$ as
constraints on incomplete inflation. The more accurately we can constrain the Universe to
be spatially flat, the stronger the evidence for an inflationary attractor with a large number of
e-foldings. 

Models have been suggested that skew inflation even more strongly to
small numbers of e-foldings \citep{Hawking:1998}. However, 
the main purpose of this example,  is to emphasise that there
is no good physical justification to adopt a uniform prior in
$\Omega_{\rm K}$ when analysing cosmological data. Since $\Omega_{\rm K}$ is
poorly determined from CMB power spectra alone (with a non-Gaussian tail
extending to large negative values) , it is
dangerous to interpret Bayesian posterior distributions in
$\Omega_{\rm K}$ as probability distributions unless one can 
justify the  choice of prior\footnote{The dangers of adopting simple priors
  in cosmology have been discussed previously by one of us
  \citep{Efstathiou:2008}.}. As a result, perceived tensions on the value of $\Omega_{\rm K}$ between
\Planck\ and other cosmological data 
are on a very different footing to tensions in the value
of, for example,  the Hubble constant $H_0$. As is well known, late time measurements
of $H_0$ differ from the base\footnote{As in the \Planck\ papers we
  refer to the six parameter \LCDM\ model (spatially flat, power law
  scalar adiabatic fluctuations, cosmological constant) as the base
  \LCDM\ model.} \LCDM\ value determined from \Planck\ by about
$4.3\sigma$ \citep[e.g.][]{Riess:2019}. However, the Hubble constant is so well
determined by \Planck\ ($H_0 = 67.44 \pm 0.58 \ {\rm km}\ {\rm s}^{-1}\ {\rm Mpc}^{-1}$)
that we can be confident that the data overwhelms the priors, since 
it is extremely unlikely that
 $H_0$ is drawn to any particular value by an attractor.

\section{Likelihoods and constraints on $\Omega_{\rm K}$}

 The \Planck\ preference for closed universes has been pointed out in
 previous \Planck\ papers \citep{Params:2014, Params:2016,
   Params:2018} and is closely related to the preference for the
 \Planck\ temperature spectra to favour more lensing than expected in
 the base \LCDM\ model (quantified by the phenomenological $A_L$
 parameter \citep{Calabrese:2008}) since both effects are caused by
 the same features in the \Planck\ temperature power spectrum in the
 multipole range $\ell \sim \mbox{1200\,--1500}$.  PCP18 also pointed out that
 when \Planck\ high multipole polarization spectra were included, the
 \plik\ TTTEEE likelihood pulls $A_L$ and $\Omega_{\rm K}$ away from
 the base \LCDM\ model with a higher significance than our own
 \camspec\ TTTEEE likelihood. The posterior for the parameter
 $\Omega_{\rm K}$ is therefore sensitive to both choice of prior and
 to the likelihood implementation.

PCP18 discussed the possibility that both the $\Omega_{\rm K}$ and
$A_L$ `tensions' were a result of statistical fluctuations.  Following
the completion of PCP18, we investigated this possibility in detail
\citep[][hereafter EG]{Efstathiou:2019} by constructing a
\Planck\ likelihood (which we refer to as the 12.5HMcln likelihood)
using more sky in temperature and polarization than in the
\Planck\ \camspec\ likelihood reported in PCP18. The construction of
the 12.5HMcln likelihood is discussed at length in EG, to which we
refer the reader for further details and for tests of the consistency
of the TE and EE polarization spectra. Increasing the sky area
reduced the `tensions' in $\Omega_{\rm K}$ and $A_L$, as expected if
they were caused by statistical fluctuations. Furthermore, we
demonstrated that the features in the temperature power spectrum that
drive these tensions are repeatable to high accuracy between the
$217\times217$, $143\times 217$ and $143\times 143$ GHz temperature
cross-spectra. It therefore seems unlikely that the $\Omega_{\rm K}$
and $A_L$ results are influenced by systematic errors in the
\Planck\ data. In addition, the polarization spectra are essentially
neutral with respect to the parameters $\Omega_{\rm K}$ and $A_L$.

\begin{figure*}
	\centering
	\includegraphics[width=58mm, angle=0]{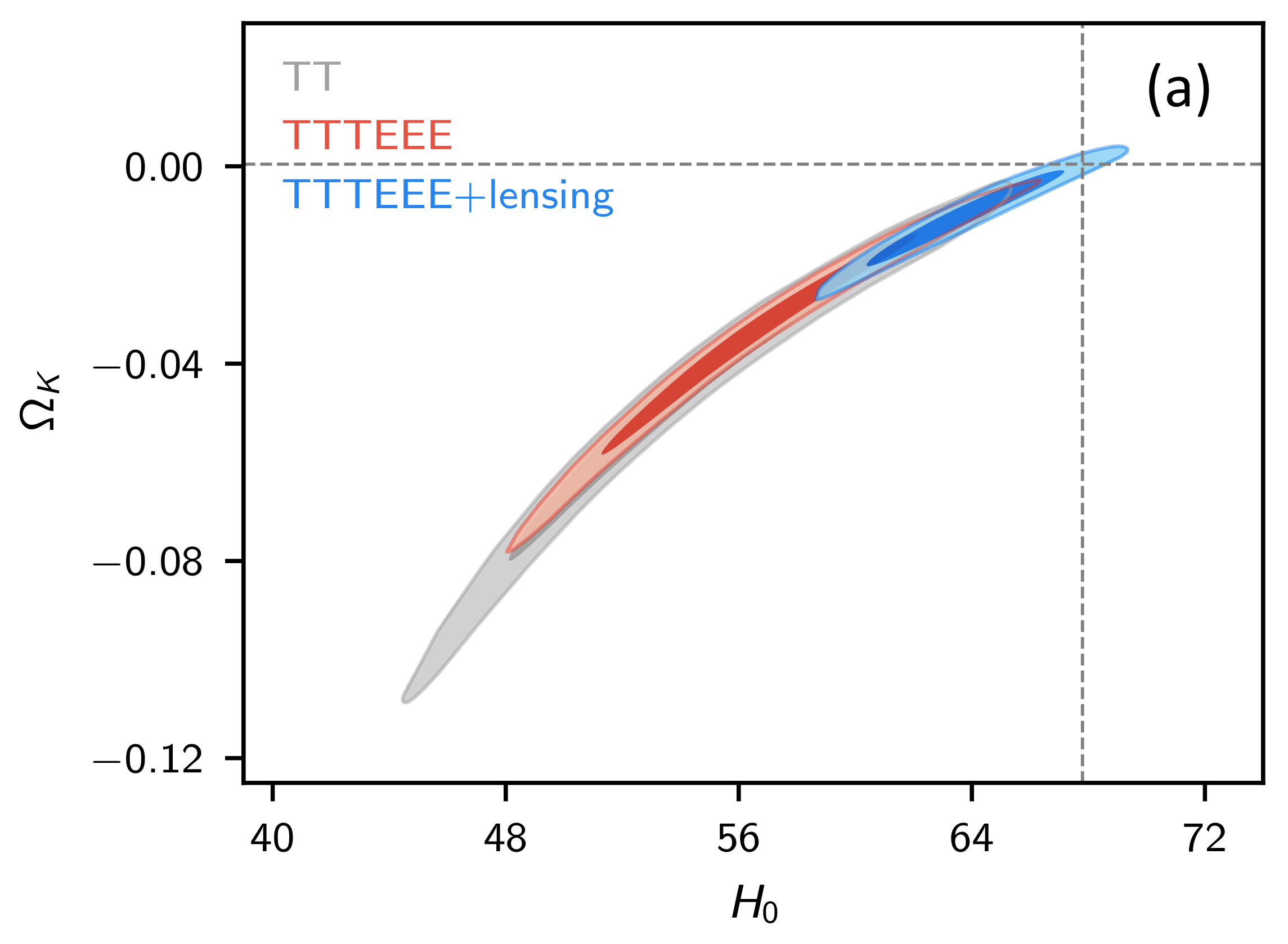} \includegraphics[width=58mm, angle=0]
{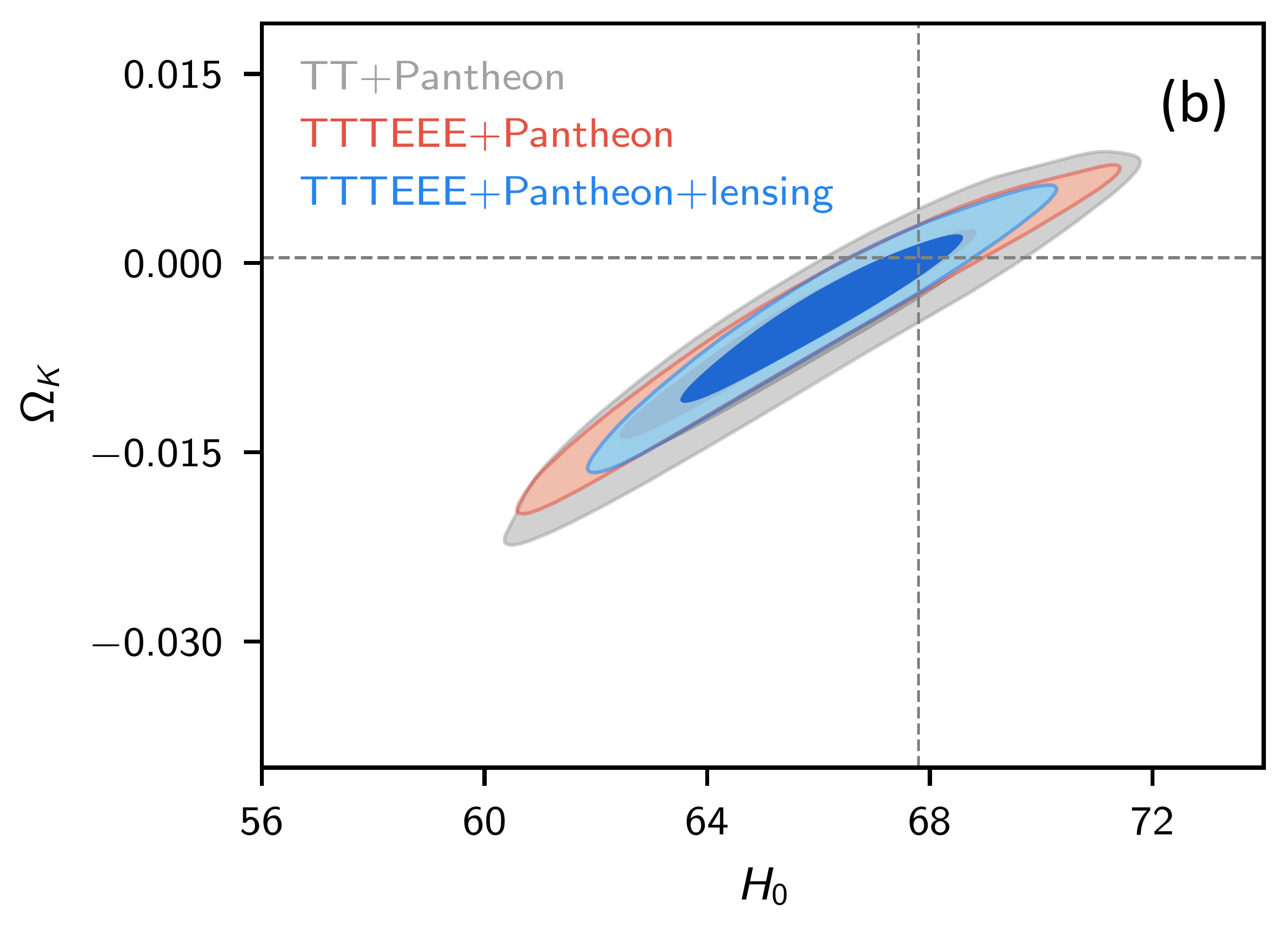}
\includegraphics[width=58mm, angle=0]{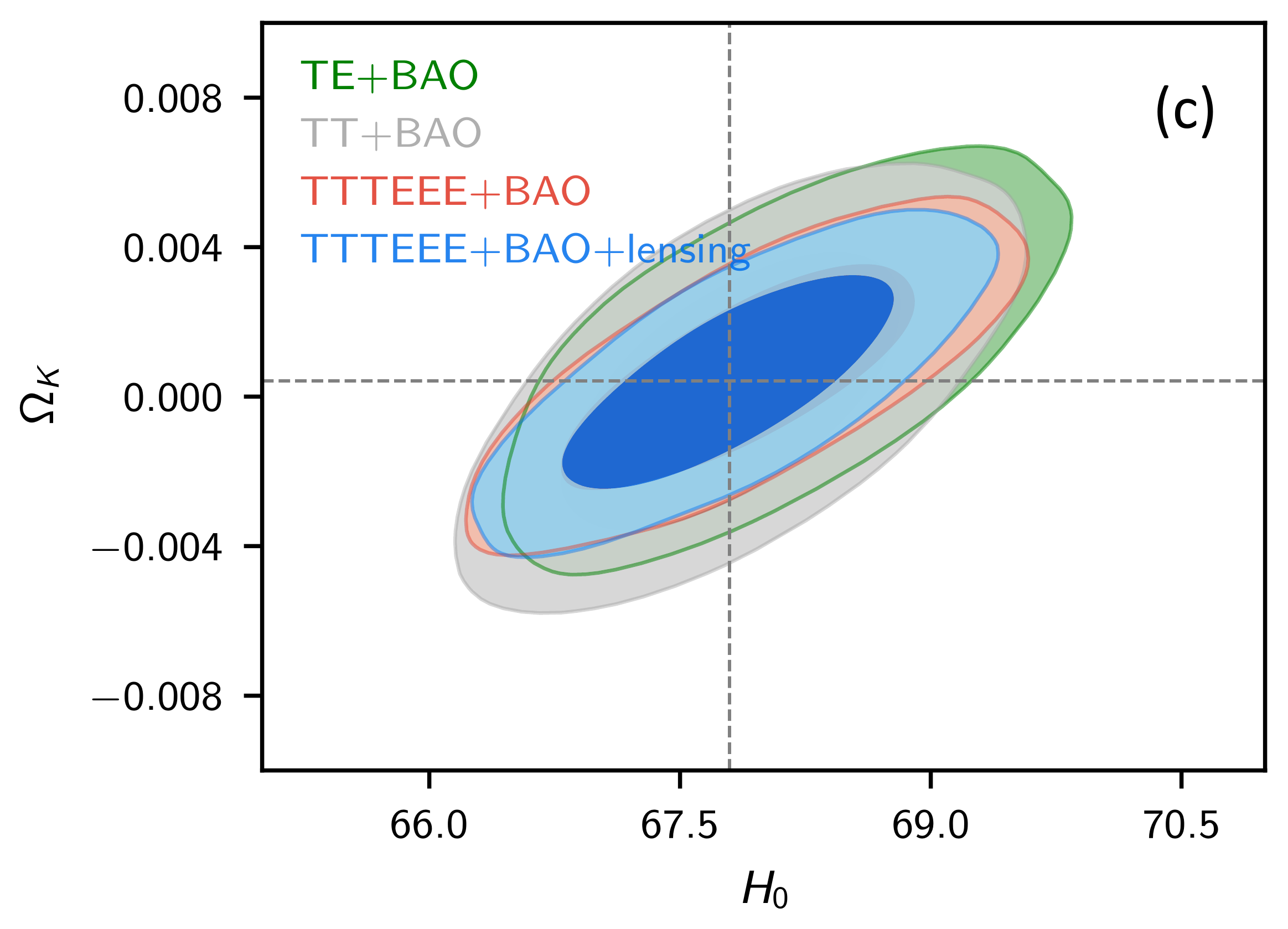}
	\caption{68\% and 95\% contours on $\Omega_{\rm K}$ and $H_0$ {\it assuming a flat prior in $\Omega_{\rm K}$ over the range $-0.3 < \Omega_{\rm K} < 0.3$}
for: (a) \Planck\ data alone; (b) \Planck\ data combined with the Pantheon supernova sample; (c) \Planck\ data combined with BAO.}

	\label{fig:omega_results}

\vspace{-0.2in}

\end{figure*}

In this paper, we use the 12.5HMcln likelihood at $\ell \ge 30$, as
described in EG, together with the $2018$ \Planck\ temperature and
polarization likelihoods at $\ell < 30$ as described in
\cite{Likelihood:2019}. The notation for these data combinations
follows that of EG, thus TTTEEE denotes the full 12.5HMcln likelihood
at $\ell \ge 30$ combining the temperature-temperature (TT),
temperature-polarization (TE) and polarization-polarization (EE)
cross-spectra; TT or TE denotes use of only the
temperature-temperature or temperature-polarization blocks of the
likelihood. The additional data are as follows: `lensing' denotes the
\Planck\ 2018 lensing likelihood as described in \cite{Lensing:2018};
`Pantheon' denotes use of the Pantheon Type Ia supernovae sample of
\citep{Scolnic:2015}; `BAO' denotes use of the baryonic acoustic
oscillation measurements from \cite{ Beutler:2011, Ross:2015,
  Alam:2017} as discussed in PCP18 and EG. In running Monte-Carlo
Markov chains\footnote{All chains were produced using {\tt COSMOMC}
  (see {\tt https://cosmologist.info/cosmomc/}) using the {\tt CAMB}
  Boltzmann code (see {https://camb.info/readme.html}) in exactly the
  same way as described in PCP18.}, we use the same priors on the six
parameters of the base \LCDM\ model as in PCP18. We treat $\Omega_{\rm
  K}$ as a one-parameter extension of the base \LCDM\ cosmology,
adopting a uniform prior over the range $-0.3 \le \Omega_{\rm K} \le
0.3$ as in PCP18 (though our interpretation of the posteriors
will differ).

With these assumptions, the posterior distributions in $\Omega_{\rm
  K}\mbox{--}H_0$ plane are shown in Fig. \ref{fig:omega_results} for various
data combinations. Fig. \ref{fig:omega_results}a shows results from
\Planck\ data alone.  By including $\Omega_{\rm K}$ as a parameter,
CMB power spectra show a strong geometrical degeneracy
\citep{Bond:1997} which is partly broken by gravitational lensing of
the CMB. The TT likelihood shows a mild pull towards negative values
of $\Omega_{\rm K}$, though with low values of $H_0$ that are strongly
disfavoured by direct measurements \citep{Riess:2019}. Adding
\Planck\ polarization spectra shifts the contours towards $\Omega_{\rm
  K} =0$, and adding \Planck\ lensing shifts the contours even closer
to the $\Omega_{\rm K}=0$ attractor
solution. Fig. \ref{fig:omega_results}b shows what happens if we
combine the \Planck\ data with the Pantheon supernovae sample. The
geometrical degeneracy is now broken and the addition of $\Omega_{\rm
  K}$ as a parameter offers very little improvement in the fits to the
\Planck+Pantheon likelihoods compared to the base \LCDM\ model.
Fig. \ref{fig:omega_results}c shows the results of adding BAO to the
\Planck\ data. The BAO data break the geometrical degeneracy very
effectively. One can see that TE+BAO likelihood combination gives very
similar constraints to the TT+BAO likelihood.  In other words, the
posteriors shown in Fig. \ref{fig:omega_results}c are insensitive to
the features at $\ell \sim \mbox{1200\,--1500}$ in the TT spectrum responsible
for the $A_L$ `tension' (see EG for further discussion). The constraints
on $\Omega_{\rm K}$ shown in Fig. \ref{fig:omega_results}c are now so
tight that they overwhelm the power law tails in a distribution such
as that of equ.  (\ref{prob2}). Marginalising over all other
parameters, we find (with notional $1\sigma$ errors given the priors):
\begin{subequations}
\begin{eqnarray}
\hspace{-0.2in} \Omega_{\rm K} = 0.0002  \pm 0.0025, &\hspace{-0.2in} &  {\rm TT+BAO}, \qquad \qquad  \qquad  \label{res1a}\\
\hspace{-0.2in}\Omega_{\rm K} = 0.0010  \pm 0.0023, &\hspace{-0.2in} &    {\rm TE+BAO},  \qquad  \label{res1b}\\
\hspace{-0.2in}\Omega_{\rm K} = 0.0005  \pm 0.0020, &\hspace{-0.2in} &   {\rm TTTEEE+BAO}, \label{res1c}\\
\hspace{-0.2in}\Omega_{\rm K} = 0.0004  \pm 0.0019, &\hspace{-0.2in} &   {\rm TTTEEE+BAO+lensing}, \label{res1d} \\
\hspace{-0.2in}\Omega_{\rm K} = 0.0004  \pm 0.0018, &\hspace{-0.2in} &   {\rm TTTEEE+BAO+Pantheon} \nonumber \\
                                                  &\hspace{-0.2in} &   \ \ \ \ \ \  \ {\rm +lensing}. \label{res1e}
\end{eqnarray}
\end{subequations}
The result in (\ref{res1e}) is essentially identical to the constraint
of equ. (\ref{Omega_constraint}) derived in PCP18. We interpret these
results as extremely strong evidence that our Universe is nearly
spatially flat. Furthermore, in the context of inflationary scenarios,
these results show that our Universe has firmly locked on to the
inflationary attractor, disfavouring models of incomplete inflation
with a limited numbers of e-foldings ($N \approx N_*$). This is a
highly non-trivial result.

\cite{DiValentino:2019} argue that observational evidence for a
closed Universe would present a crisis for cosmology.  We agree with
this conclusion. If the Universe were indeed closed with a value of
$\Omega_{\rm K} \approx -0.04$ then one would have to argue that
unexpected  new physics or systematics  in the \Planck\ lensing data, supernovae and BAO
all act in the same way to favour $\Omega_{\rm K}=0$.
Since these data sets are independent of each other and respond to different physics (supernovae and BAO test the background cosmology, while lensing
tests theory at the level of perturbations)  this is 
extraordinarily unlikely\footnote{Note also that CMB lensing measurements from the South Pole Telescope
agree well with the \Planck\ lensing measurements and strongly favour a spatially flat Universe when combined with 
\Planck\ \citep{Bianchini:2020}.}. It is much more plausible that these additional datasets
break the geometrical degeneracy leading to values of $\Omega_{\rm K}$ that are closer
to the truth. The fact that all three datasets favour $\Omega_{\rm K} = 0$ provides powerful
evidence that our Universe is nearly spatially flat. 

\begin{table*}

\label{tab:chi_squared}
\begin{center}

\caption{$\chi^2$ values for best fit cosmologies with and without curvature.}
\small

\smallskip

\begin{tabular}{l|c|c|c|c|c|c|} \hline 
  & \multicolumn{3}{|c|} {{\rm TT}}& \multicolumn{3}{|c|} {{\rm TTTEEE}}\\ 
(fits with lowl) &   base  & $\Omega_{\rm K}$ & & base  & $\Omega_{\rm K}$ &  \\
Likelihood     &   $\chi^2_{\rm min}$ &  $\chi^2_{\rm min}$ &  $\Delta \chi^2_{\rm min}$ &  $\chi^2_{\rm min}$ &  $\chi^2_{\rm min}$ &  $\Delta \chi^2_{\rm min}$ \\ \hline
lowl TT ($\ell < 30$)   &$\ 23.73$  & $\ 21.13$   & $-2.60$  &  $\ 23.864$  & $\ 21.38$ & $-2.30$       \\
lowE EE ($\ell < 30$)   &$396.35$  & $395.57$   & $-0.78$  &  $396.07$  & $395.60$ & $-0.47$       \\
\camspec\ ($\ell \ge 30$)   &$5491.42$  & $5489.30$   & $-2.13$  &  $9790.96$  & $9787.93$ & $-3.02$       \\ \hline
(fits without lowl) &   base  & $\Omega_{\rm K}$ & & base  & $\Omega_{\rm K}$ &  \\
Likelihood     &   $\chi^2_{\rm min}$ &  $\chi^2_{\rm min}$ &  $\Delta \chi^2_{\rm min}$ &  $\chi^2_{\rm min}$ &  $\chi^2_{\rm min}$ &  $\Delta \chi^2_{\rm min}$ \\ \hline
lowE EE ($\ell < 30$)   &$395.57$  & $395.62$   & $+0.08$  &  $395.77$  & $395.62$ & $-0.15$       \\
\camspec\ ($\ell \ge 30$)   &$5491.02$  & $5489.47$   & $-1.82$  &  $9791.23$  & $9787.91$ & $-3.32$       \\ \hline

\end{tabular}
\end{center}
\end{table*}

Another possibility is that the tendency for \Planck\ power spectra to
favour closed Universes is caused by systematic errors in the
\Planck\ likelihoods and/or \Planck\ data. As discussed above, it is
certainly true that different likelihood implementations lead to
different results, with the \plik\ likelihood favouring closed
Universes more strongly than our own \camspec\ likelihood.  We have
discussed the construction of the \camspec\ likelihood in great detail
in EG and have argued that our methodology is robust and gives
reasonable $\chi^2$ values for the polarization spectra, unlike
\plik\ \citep[see][for further details]{Likelihood:2019}. However, for
readers interested in spatial curvature, whether \plik\ or
\camspec\ is the more reliable likelihood is irrelevant because {\it
  differences between} \Planck\ {\it likelihoods} {\it are overwhelmed
  when} \Planck\ {\it data are combined with BAO}. This is why the
estimates of equs. (\ref{Omega_constraint}) and (\ref{res1d}) agree so
precisely.

The final question to consider is whether there is a statistical
inconsistency, i.e. if we allow $\Omega_{\rm K}$ to vary, are the fits
to the \Planck\ power spectra so much better than the fits to the base
\LCDM\ model to suggest systematics or new physics? We have already
argued that the posterior distributions for $\Omega_{\rm K}$ should
not be interpreted as probability distributions because of their
sensitivity to priors. Likewise, evidence ratios can give misleading results
because of sensitivity to priors \citep{Efstathiou:2008}.  
Since the models are nested, we can answer this
question in a definitive and particularly simple way, independent of priors, by looking at
differences in $\chi^2$ values, i.e. likelihood ratios\footnote{Other statistical measures for model
  selection are discussed by {e.g.} \cite{Liddle:2004, Liddle:2007,
    Handley:2019b}.}. Table \ref{tab:chi_squared} lists values of
$\chi^2 = - 2\ln{\cal L}$ for the best fit cosmology for the base
\LCDM\ cosmology with $\Omega_{\rm K} = 0$ and for the best fit when
$\Omega_{\rm K}$ is allowed to vary as an additional parameter. We
have decomposed the likelihood into the various components: the
\commander\ temperature likelihood at $\ell < 30$ (denoted `lowl'),
the \simall\ polarization likelihood at $\ell < 30$ (denoted `lowE')
and the \camspec\ likelihood at $\ell \ge 30$.

 The overall reduction in $\chi^2$ is about $6$, split roughly equally
 between the lowl likelihood and \camspec. (The lowE likelihood is
 neutral to the addition of $\Omega_{\rm K}$.) Adding $\Omega_{\rm K}$
 as a parameter reduces the \camspec\ $\chi^2$ values by $2.13$ (TT)
 and $3.02$ (TTTEEE). These are very modest changes and are consistent
 with the conclusion of EG that the base \LCDM\ model provides
 essentially a perfect fit to the \Planck\ power spectra at $\ell \ge
 30$ as judged by $\chi^2$ statistics. The improvement in the fits to
 the low multipole likelihood is a consequence of the low 
 amplitudes of the low multipoles (including the quadrupole) relative to the predictions of the
 base \LCDM\ model noted in previous \Planck\ papers \citep[see
   e.g.][]{Params:2014}. There is, however, an additional subtelty
 involved in interpreting the low multipoles.  In {\tt CAMB} the power
 spectrum in non-flat models is written as
\begin{equation}
 P(k) = {(q^2 - 4K)^2 \over q(q^2 - K)} k^{(n_s-1)},  \label{PS}
\end{equation}
where $q = \sqrt{k^2 + K}$, which is a highly specific assumption on how the fluctuation spectrum extends to scales
greater than the curvature scale. This form leads to a suppression of the low multipoles in closed models \citep[see][]{Efstathiou:2003} and can
compensate with other parameters to reduce the $\chi^2$ of the lowl likelihood. Since equ (\ref{PS}) is not based on
any specific theory, we should not assign much weight to the reduction in $\chi^2$ in the lowl likelihood. A more
reasonable statistical approach would be to add additional parameters to describe the fluctuations on scales
greater than the curvature scale. If we exclude the lowl likelihood entirely, the best fits to \camspec+lowE 
are shifted slightly towards spatially flat universes with minimum $\chi^2$ values as listed in the last
two lines of Table \ref{tab:chi_squared}. The overall shifts in $\chi^2$ in \camspec\ are small and very similar
to those when lowl is included. The fits to the high multipole data from \Planck\ therefore are barely improved if 
curvature is added as an additional parameter to base \LCDM. 

Since many researchers are more comfortable with ptes than likelihood ratios, 
we can translate as follows \citep{wilks}. Assume that $\Delta \chi^2_{min}$ is drawn from a  $\chi^2$ distribution with
$1$ degree of freedom,  the total change in $\chi^2$ from low and high
multipoles suggests that the \Planck\ data, excluding \Planck\ lensing,
favour $\Omega_{\rm K} < 0$ with a pte of about  $1.6 \%$  (or a Gaussian $2.1 \sigma$). However, 
note the caveat above concerning low multipoles in temperature. If we exclude the TT spectrum at $\ell < 30$, the pte rises to 
about $7\%$  (or a Gaussian $1.6\sigma$). These numbers are very different from the pte of $0.034\%$ ($3.4 \sigma$)
 quoted by \cite{DiValentino:2019}.

\section{Conclusions}

The geometry of the Universe is a question of fundamental importance
to cosmology.  We have argued that the claims in
\cite{DiValentino:2019} that \Planck\ data strongly favour closed Universes at
high significance are a consequence of using the \plik\ TTTEEE likelihood
which differs from the \camspec\ likelihood and 
ignoring the importance of
priors.  There is no good reason to assume a uniform prior on
$\Omega_{\rm K}$ and so the posterior for $\Omega_{\rm K}$ and ptes
derived from it should not be over-interpreted.  We have presented
results from a new \Planck\ likelihood that shows a weak and
statistically insignificant pull towards closed universes. This
tendency is overwhelmed when the \Planck\ likelihood is combined with
other types of data that break the geometrical degeneracy.  Combining
\Planck\ power spectrum measurements with any one of \Planck\ CMB
lensing, Type Ia supernovae or BAO data, favours a spatially flat
universe. The strongest constraint (equ. \ref{res1e}) shows that the
Universe is spatially flat to a precision of $\sim 0.0018$, in agreement with the results in PCP18.  This is a
profound result for inflationary cosmology. If inflation is indeed the
solution to the flatness problem, the observations show that the
Universe must have firmly locked on to the $\Omega_{\rm K}=0$
attractor. Models of incomplete inflation, with e-foldings $N\sim
N_*$, are disfavoured by observations.

\section*{Acknowledgements} We thank Anthony Challinor for useful discussions. 
Steven Gratton acknowledges the award of a Kavli Institute Senior Fellowship.
We are grateful to Antony Lewis for maintaining the {\tt COSMOMC} software used in this paper
and for comments on the manuscript.

\bibliographystyle{mnras}
\bibliography{omega} 
\end{document}